\documentclass[useAMS,usenatbib,a4paper]{mn2e}

\usepackage{amssymb,amsmath}
\usepackage{graphicx}
\usepackage{natbib}
\usepackage{aas_macros}
\usepackage{color}

\newcommand{\hii}{H~{\sc ii}}

\usepackage{soul}

\begin{document}

\title[Morphologies of protostellar outflows]{Morphologies of protostellar outflows: An ALMA view}

\author[Thomas Peters et al.]{
\parbox[h]{\textwidth}{
Thomas Peters$^1$\thanks{E-Mail: tpeters@physik.uzh.ch},
Pamela D. Klaassen$^2$,
Daniel Seifried$^3$,
Robi Banerjee$^3$\\
and Ralf S. Klessen$^4$}\vspace{0.4cm}\\
\parbox{\textwidth}{$^1$Institut f\"{u}r Theoretische Physik, Universit\"{a}t Z\"{u}rich,
Winterthurerstrasse 190, CH-8057 Z\"{u}rich, Switzerland\\
$^2$Leiden Observatory, Leiden University, PO Box 9513, 2300 RA Leiden, The Netherlands\\
$^3$Hamburger Sternwarte, Gojenbergsweg 112, D-21029 Hamburg, Germany\\
$^4$Universit\"{a}t Heidelberg, Zentrum f\"{u}r Astronomie, Institut f\"{u}r Theoretische Astrophysik, Albert-Ueberle-Str. 2, D-69120 Heidelberg, Germany}}

\maketitle

\begin{abstract}
The formation of stars is usually accompanied by the launching of protostellar outflows. Observations
with the Atacama Large Millimetre/sub-millimetre Array (ALMA) will soon revolutionalise
our understanding of the morphologies and kinematics of these objects. In this paper, we
present synthetic ALMA observations of protostellar outflows based on numerical magnetohydrodynamic
collapse simulations. We find significant velocity gradients in our outflow models
and a very prominent helical structure within the outflows. We speculate that the disk wind
found in the ALMA Science Verification Data of HD~163296 presents a first instance of such
an observation.
\end{abstract}

\section{Introduction}

Protostellar outflows are generally byproducts of star formation in the full range
from low- to high-mass star-forming regions
\citep{caband91,bachiller96,reibal01,shepherd05,beuthshep05,arceetal07,baletal07,bally07,bally08}.
Here we focus our attention on outflows from intermediate-mass protostars of a few solar masses,
with typical mass-loss rates of $10^{-5}$ to a few $10^{-3}\,M_\odot\,$yr$^{-1}$
\citep{beutheretal02,zhangetal05,renetal11,wangetal11} and
outflow momentum rates from $10^{-4}$ to several $10^{-2}\,M_\odot\,$km\,s$^{-1}$\,yr$^{-1}$
\citep{beutheretal02,zhangetal05,shietal10,wangetal11}.
Intermediate-mass outflows are typically elongated with collimation factors between
1 and 10 \citep{ridmoo01,wuetal04,beutheretal02b,beuthetal04a}, but
recent observations in W75N have revealed an apparently very young, spherical outflow
\citep{torrelles2003,surcisetal11,kimkim2013}.

There are two essentially independent mechanisms that can drive protostellar outflows with the help of
magnetic fields. First, the disk material can be accelerated centrifugally
and launch a disk wind \citep[e.g.][]{blandfordpayne82,pudrnorm83,pelput92}. Or second, the gas in the disk can be lifted by the pressure of the toroidal magnetic field
in a magnetic tower flow \citep[e.g.][]{lynbel96,lynbel03}. Based upon the Lorentz force, these mechanisms are not mutually exclusive and can well act in concert.
\citet{seifried12} derived an analytic description able to distinguish both
types of driving mechanisms along the outflow region which were also
tested with numerical simulations.
Additionally, outflows around high-mass stars can be driven by radiation pressure \citep{krumholzetal09,kuiperetal12} and ionization feedback \citep{petersetal10a,petersetal12,klaassen2013b}.
However, these types of outflows have a very different morphology and kinematics that are on the lower end of observed outflow properties \citep{petersetal12,klaassen2013b}.

Three-dimensional magnetohydrodynamic models of protostellar outflows
\citep{seifetal11,seifried12} show that magnetically-driven outflows
can have strongly varying morphologies, ranging from collimated, elongated
outflows to almost spherical bubbles. \citet{seifried12} found that
collimated outflows are only formed when a nearly
Keplerian disk is present. In idealized numerical simulations of protostellar collapse
(i.e. simulations without initial turbulence) Keplerian disks build up
only in cases of weak magnetic fields \citep{hentey08,hencia09}, whereas Keplerian disks
naturally arise in simulations that include some initial velocity or
density perturbations independent of the magnetic field strength \citep{seifried12a,seifried13,santoslima12,myers13}. The morphologies of the outflows launched from
those turbulence-generated disks are not yet studied in detail. It
seems likely that that the associated outflows will be launched by
magnetic pressure gradients as well as by magneto-centrifugal forces
and will take on different shapes depending on the environment and
evolutionary state of the underlying disk.

In more massive initial cloud cores, \citet{petersetal11a} found that magnetic bubbles can also be
produced from common disks around multiple systems
by fragmentation-induced outflow disruption. 
Here, spherical, low-velocity outflows are generated by a large-scale pseudodisk or toroid that forms around the central high-mass star,
which then becomes gravitationally unstable and fragments. This fragmentation process destroys any coherent rotation
and renders both magnetic launching mechanisms impossible. Hence, as more gas falls onto the disk,
the radius at which gravitational instability sets in increases and the outflow stalls inside of this radius. Since the
toroid itself grows in radius as well and the outer parts of the disk still rotate coherently, the outflow
becomes more spherical in shape. \citet{giratetal2013} have recently reported the discovery of such a large-scale spherical
tower flow. Fragmentation-induced outflow disruption thus naturally relates to the idea
of fragmentation-induced starvation
\citep{petersetal10a,petersetal10c,girietal12}.
Even if individual outflows are launched from the disks around single
low- and high-mass protostars \citep[which are not resolved in the simulations by][]{petersetal10a} within the young multiple system, it
might be difficult to observe collimated outflows due
to the mutal interaction of those outflows and the influence of the
\hii\ regions around the massive protostars. 

In all cases, the
outflows found in numerical simulations show a complex internal
structure where knots are generated and different instabilities
occur. Often prominent are corkscrew and helical structures due to
kink instability or outflow precession \citep[e.g.][]{ouyclapud03,staff10}. Although those internal structures are commonly observed
in jets from low-mass stars \citep[see e.g., review by][]{rayetal07},
the outflow structure from intermediate- and high-mass stars is more
obscure. Yet, the recent observation of an outflow from the young
A-type star HD 163296 by \citet{klaassen2013} reveals a double sided
corkscrew structure that is interpreted as the internal structure of a
disk wind launched from around this star. In this present study we
confront those observations with numerical simulations. 

To avoid the complexities of feedback and fragmentation in high-mass star formation, we here focus on outflows from intermediate-mass stars.
We make predictions for how the different outflow
morphologies found by \citet{seifried12} would appear to the observer,
in particular focusing on the recent observations by \citet{klaassen2013}.
In Section~\ref{sec:RTmodels}, we describe the simulation snapshots that were observed following the procedure
outlined in Section~\ref{sec:synobs}. We present the results of our analysis in Section~\ref{sec:res}
and compare our findings with observations in Section~\ref{sec:compobs}. We conclude in Section~\ref{sec:concl}.

\section{Description of the simulation snapshots}
\label{sec:RTmodels}

The simulations are performed with the astrophysical code FLASH~\citep{fryxell00} using the MHD solver devised by
\citet{bouchut07}. To follow the long-term evolution of the protostellar disks and their associated
outflows we make use of sink particles~\citep{federrathetal10}. For details of the numerical methods applied we refer to~\citet{seifetal11}.
Since these simulations to not include radiative feedback, the gas can only heat up through hydrodynamic
processes, such as compression and the development of shocks.

In the following we analyse the results of two simulations which differ only in the strength of the initial magnetic field.
Both simulations start with a 100~$M_{\sun}$ molecular cloud core, 0.25~pc in diameter and rotating rigidly around the $z$-axis with
a rotation frequency of $3.16 \cdot 10^{-13}$ s$^{-1}$. The magnetic field is initially aligned with the rotation axis, i.e. parallel
to the $z$-axis. In run 26-4 the magnetic field strength is chosen such that the normalized mass-to-flux ratio is
\begin{equation*}
  \mu = \left(\frac{\rmn{M_{core}}}{\Phi_{\rmn{core}}}\right)/\left(\frac{M}{\Phi}\right)_{\rmn{crit}} = \left(\frac{\rmn{M_{core}}}{\int B_z \mathrm{d}A}\right)/\left(\frac{0.13}{\sqrt{G}}\right)
= 26,
\end{equation*}
and the ratio of rotational to gravitational energy is $\beta_\mathrm{rot} = 4 \cdot 10^{-2}$ (the run number encodes
these two fundamental quantities of the initial conditions).
Hence, the core is magnetically supercritical and the magnetic field does not have a strong impact on the collapse of the core.
In the second run 5.2-4, the magnetic field strength is increased by a factor of 5 resulting in a mass-to-flux ratio $\mu$ of
5.2. Again, we refer to \citet{seifetal11} for more details on the initial conditions.

During the collapse of the core in run 26-4, a rotationally supported disk builds up around the first protostar, starting to
fragment after $\sim$ 2600~yr. In contrast, in run 5.2-4 a sub-Keplerian disk with strong radial infall motions forms, showing
no signs of fragmentation until the end of the simulation. This is a consequence of the efficient removal of angular momentum
from the inner parts of the core by magnetic braking~\citep{moupal80}. On the other hand, for run 26-4 magnetic braking
is too weak so that a Keplerian disk can build up.

In both simulations a magneto-centrifugally driven protostellar outflow is launched after the formation of the first sink particle.
The outflow in run 26-4 has a well-collimated morphology with a collimation factor of $\sim$ 4 by the end of the simulation. The outflowing
gas reaches velocities of up to $\sim$20 km s$^{-1}$, well above the escape speed. In contrast, in run 5.2-4 a poorly collimated, almost spherical
outflow with relatively low outflow velocities up to $\sim$7 km s$^{-1}$ is formed. Both outflows keep expanding in a roughly self-similar fashion keeping
the overall morphological properties.

While only a single sink particle forms in run 5.2-4, a small cluster develops in run 26-4. The
mass spectrum in this cluster is, however, by far dominated by the central sink particle in our snapshots.

We analyse three snapshots in total, two snapshots for run 26-4 (Elon-A and Elon-B) and one snapshot of run 5.2-4 (Spher).
The masses of the central sink particles in these snapshots are 2.02\,$M_\odot$, 2.89\,$M_\odot$ and 2.29 \,$M_\odot$, respectively.
Table~\ref{table:direct} shows the mass, momentum and kinetic energy of the outflows. The values are
determined directly from the simulation data by measuring all the outflowing gas more than 50~AU
above and below the midplane. Unlike observational measurements, we do not restrict contributions to the summed values to a certain
velocity range along a particular line of sight.

The outflow Elon-A of run 26-4 has an average outflow lobe height of 3200~AU and an age of 5000~yr, but during the first
1500~yr the outflow grows only very slowly. The volume-weighted mean temperature is 60~K (41~K mass-weighted), averaged spatially over the outflow region.
These larger values compared to the initial temperature (20~K) are primarily due to compressive motion and shock heating.
The maximum outflow speed is 19~km~s$^{-1}$. Note that this maximal velocity is attained close to the disk and that
most of the gas in the outflow is a few km~s$^{-1}$ slower.

In snapshot Elon-B of run 26-4, the maximal velocity is only 14~km~s$^{-1}$ because
disk fragmentation at 5500--6000~yr reduces the outflow activity. The outflow is now 10000~yr old (or 8500~yr,
disregarding the initial slow starting phase) and has an average outflow lobe height of 9600~AU, consistent
with a linear extrapolation from the previous snapshot with a typical outflow velocity of 14~km~s$^{-1}$.
By comparing the masses and kinematics of Elon-A and the later time Elon-B in
Table~\ref{table:direct}, we find that the momentum and energy of the outflow do not grow by the same
factor as the outflow mass. In fact, the outflow energy even decreases slightly for the blue-shifted component.
The volume-weighted mean temperature is a bit smaller than before and now has the value 52~K (39~K mass-weighted).
All of these effects are results of the disk fragmentation. 

Snapshot Spher of run 5.2-4, with an average height of one outflow lobe of 1100~AU, is 4000~yr old. The maximum outflow speed is
6.9~km~s$^{-1}$, less than half the velocity observed in the other simulation. The volume-weighted mean temperature
is 33~K (28~K mass-weighted). It is expected that the gas temperature here is lower compared to the other simulations
because the smaller outflow velocities compress the gas to a lesser extent. Run~5.2-4 has not been followed for longer times,
so that we cannot analyse another snapshot from a later stage.

\begin{table*}
\caption{Outflow parameters determined from the simulations}
\label{table:direct}
\begin{center}
\begin{tabular}{lrccccc}
\hline
&& \multicolumn{1}{c}{$M$} & \multicolumn{1}{c}{$P$} & \multicolumn{1}{c}{$E$} & \multicolumn{1}{c}{$L$} & \multicolumn{1}{c}{$\dot{M}$}\\
&& \multicolumn{1}{c}{($M_\odot$)} & \multicolumn{1}{c}{($M_\odot$ km s$^{-1}$)} & \multicolumn{1}{c}{(10$^{43}$ erg)}
& \multicolumn{1}{c}{($L_\odot$)} & \multicolumn{1}{c}{($10^{-3}\,M_\odot\,$yr$^{-1}$)}\\ \hline
Elon-A &
blue& 0.46 & 1.34 & 6.65 & 0.11 & 0.092\\
&red& 0.38 & 1.02 & 5.48 & 0.09 & 0.076\\
\hline
Elon-B  &
blue& 0.76 & 1.59 & 6.12 & 0.05 & 0.076\\
&red& 0.73 & 1.42 & 6.11 & 0.05 & 0.073\\
\hline
Spher &
blue& 0.32 & 0.35 & 0.56 & 0.01 & 0.080\\
&red& 0.33 & 0.37 & 0.61 & 0.01 & 0.083\\
\hline
\end{tabular}
\medskip\\
Outflow mass $M$, momentum $P$, kinetic energy $E$, luminosity $L$ and mass-loss rate $\dot{M}$ as determined directly from the simulation data.
\end{center}
\end{table*}

\begin{table*}
\caption{Outflow parameters determined from synthetic CO observations}
\label{table:co}
\begin{center}
\begin{tabular}{lrcccccc}
\hline
&& \multicolumn{1}{c}{$M$} & \multicolumn{1}{c}{$P$}& \multicolumn{1}{c}{$E$} & \multicolumn{1}{c}{$L$} & \multicolumn{1}{c}{$\dot{M}$} \\
&& \multicolumn{1}{c}{($M_\odot$)} & \multicolumn{1}{c}{($M_\odot$ km s$^{-1}$)} & \multicolumn{1}{c}{(10$^{43}$ erg)} & \multicolumn{1}{c}{($L_\odot$)} & \multicolumn{1}{c}{($10^{-3}\,M_\odot\,$yr$^{-1}$)}\\ \hline
Elon-A &
blue & 0.59 & 2.40 & 11.2 & 0.37 & 0.24\\ 
&red & 0.45 & 2.16 & 12.3 & 0.41 & 0.18\\ 
\hline
Elon-B &
blue& 1.69 & 5.31 & 19.5 & 0.32 & 0.34\\ 
&red& 0.59 & 2.68 & 13.4 & 0.22 & 0.12\\ 
\hline
Spher &
blue& 0.28 & 0.64 & 1.63 & 0.17 & 0.34\\
&red& 0.11 & 0.32 & 0.95 & 0.10 & 0.14\\
\hline
\end{tabular}
\medskip\\
Outflow mass $M$, momentum $P$, kinetic energy $E$, luminosity $L$ and mass-loss rate $\dot{M}$  as determined from the synthetic CO observations.
\end{center}
\end{table*}

\section{Synthetic observations}
\label{sec:synobs}

We use the three-dimensional adaptive-mesh radiative transfer code
RADMC-3D\footnote{http://www.ita.uni-heidelberg.de/$\sim$dullemond/software/radmc-3d/}
to make synthetic CO line observations. We model the molecular line emission of the $J=2-1$ transition
of the isotopologues ${}^{12}$CO, ${}^{13}$CO and C$^{18}$O with abundances relative to H$_2$
of $10^{-4}$, $1.3 \times 10^{-6}$ and $1.8 \times 10^{-7}$, respectively \citep{wilroo94}.
The critical density of the CO $J=2-1$ transition is $\approx 2 \times 10^4\,$cm$^{-3}$.
Figure 2 and 7 of \citet{seifried12} show that the minimum density in the outflow is more than two
orders of magnitude higher than this value.
Because of the high density of the outflow material we can assume local thermodynamic equilibrium (LTE).
The Einstein coefficients of the transitions were taken from the Leiden
Atomic and Molecular Database \citep{schoeieretal05}.

The outputs from RADMC-3D were converted into skymaps assuming a distance to source
of 128~pc for setting the angular scale and converting the fluxes to Jy beam$^{-1}$.
Noiseless ALMA observations were simulated using the CASA tasks ``simobserve'' and
``simanalyze''. Simulations were carried out in CASA version 4.0.0 \citep{mcmullin07}.

We simulated full ALMA observations of the $J=2-1$ transition of $^{12}$CO, $^{13}$CO and C$^{18}$O with a
spatial resolution of 0.68$\times$0.59$''$, or 83~AU. We have chosen this beam size because
it is an average value for the full ALMA configurations.
 The outflows are observed at an inclination of
30$^\circ$ with respect to the disk normal direction, which is a representative outflow orientation \citep{cabber86}.
The resulting images had spectral
resolutions of 390~kHz ($\sim$ 0.5~km~s$^{-1}$), with the data cubes for the three isotopologues
centered at 230.538, 220.399 and 219.560 GHz respectively. The simulated fields of view fit
within a single ALMA Band~6 pointing, and we used a total integration time of 4~hours. Cleaning
was done non-interactively, using natural weighting, and a threshold of 18~mJy. 
This limit was chosen to best represent the emission in the maps, as we did not simulate atmospheric noise.

We calculated outflow properties from the $^{12}$CO map, after having corrected for the opacity of the
line using the less optically thick $^{13}$CO and C$^{18}$O isotopologues.  We used Equation~1 of \citet{choi93} to
derive the optical depth of $^{12}$CO using the least abundant isotoplogue (C$^{18}$O) towards the line center,
and $^{13}$CO at higher velocities when the C$^{18}$O emission fell below our threshold of 18~mJy.

To determine the gas mass in each velocity bin, we first calculated the column density from the flux in the
given velocity bin scaled by the average line opacity at that velocity,
\begin{equation}
N = \frac{Z \cdot F}{X}\frac{ \tau}{1-\exp(-\tau)},
\label{eq:tau}
\end{equation}
where $N$ is the column density of H$_2$ and $Z$ is the partition function for converting the column density in the $J$=2 level to the overall level
populations assuming LTE and using a temperature of 50~K (a compromise between the temperatures derived above in Section~\ref{sec:RTmodels}),
$F$ is the integrated intensity of the line (in units of K~km~s$^{-1}$),
and $X$ is the abundance of $^{12}$CO with respect to H$_2$.  
The gas mass in each channel was then derived by multiplying the column density by the size of the emitting region to obtain
the total number of H$_2$ molecules. This was then multiplied by the mass of hydrogen, and the mean molecular weight of 2.3.
These individual masses were summed for all velocities greater than $\pm$2 km~s$^{-1}$, -2.5,+2 km~s$^{-1}$, and  $\pm$2.0 km~s$^{-1}$ for
the three simulations (Elon-A, Elon-B and Spher, respectively) to find the total mass in each outflow lobe.
These velocities were chosen based on visual inspection of the processed spectra. The limits were
chosen individually for each simulation, and were set where the spectra first appear Gaussian.

To derive the kinematics of the outflows, we multiplied the gas mass in each channel by the velocity of that channel to derive the outflow
momentum ($P = \sum_i m_iv_i$) and mechanical energies in the outflows ($E = 1/2 \sum_i m_i v_i^2$). The outflow
luminosities and mass-loss rates were obtained by dividing the mechanical energies and masses, respectively,
by the kinematically derived ages of the outflows. Using the spatial extents of the outflows, and the
mass-weighted velocities at the end of the outflows, we determined outflow ages of 2400, 5000 and 800 yr for Elon-A,
Elon-B and Spher. We note that, especially for Spher, this is an underestimate of the true outflow age as we are
looking down the outflow cavity instead of across it in the plane of the sky.

The derived outflow masses and kinematics are presented in Table~\ref{table:co}. We note that atmospheric noise is not included in our images.
The uncertainty in our simulated observation comes from the uv coverage and integration time only. We find typical
errors of the order $10^{-4}\,M_\odot$ for $M$, $10^{-4}\,M_\odot$~km~s$^{-1}$ for $P$, $10^{40}\,$erg for $E$,
$0.1\,L_\odot$ for $L$ and $10^{-5}\,M_\odot\,$yr$^{-1}$ for $\dot{M}$. These quantities reflect the noise levels
in the synthetic observations which have been propogated through in quadrature. 
We used a beam uncertainty of 0.25$''$ and velocity uncertainty of 0.25~km~s$^{-1}$, which represent one third of our
spatial and one half of our spectral resolution, respectively. If we were to take
atmospheric noise into account, the ALMA Sensitivity Calculator expects we would have a noise level of 6 mJy beam$^{-1}$,
instead of the measured root-mean-square noise level of 1.3 mJy beam$^{-1}$. When cleaning the data, we used a threshold based on 6 mJy beam$^{-1}$.

\section{Results}
\label{sec:res}

\begin{figure*}
\includegraphics[width=210pt]{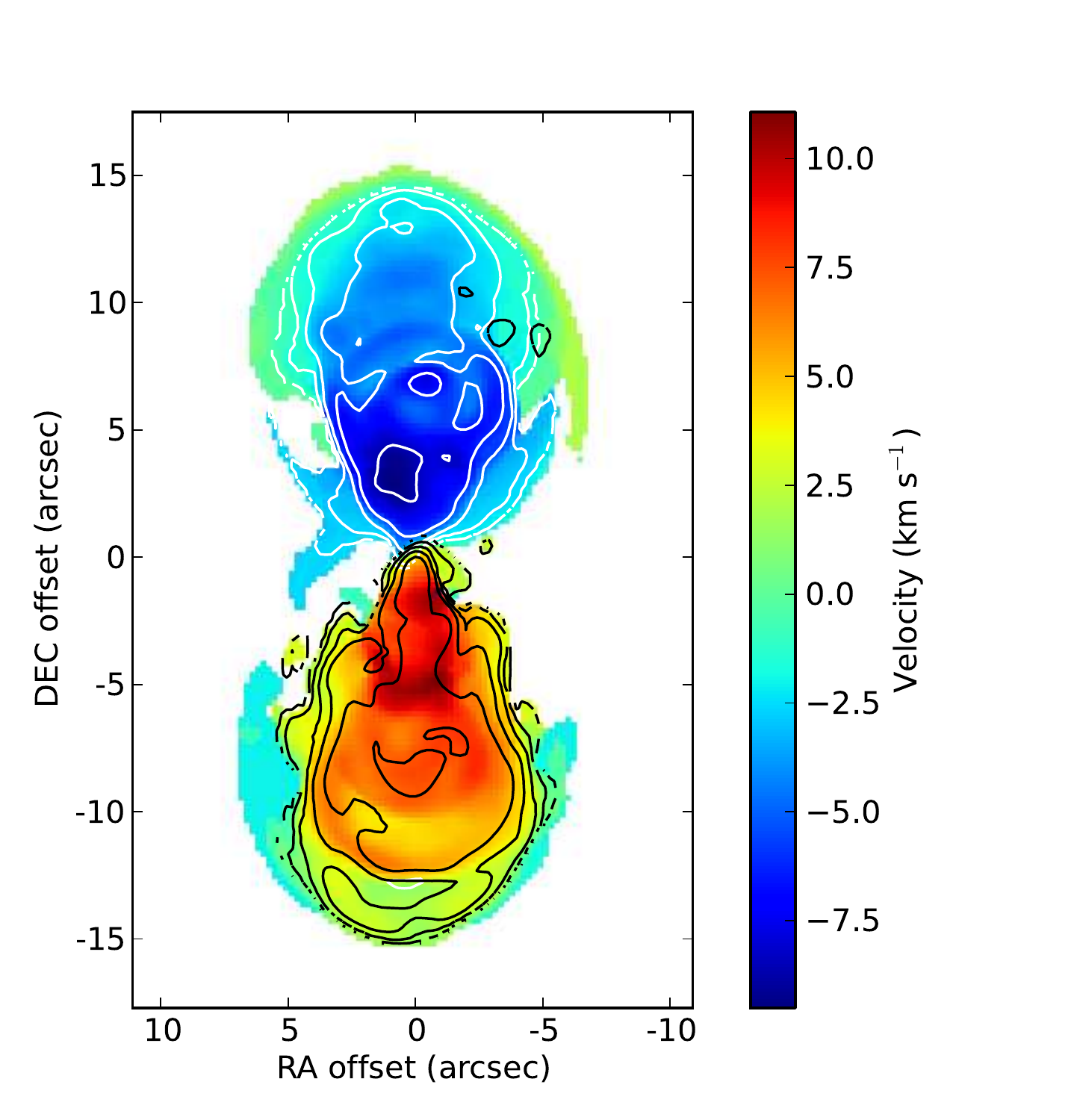}
\includegraphics[width=290pt]{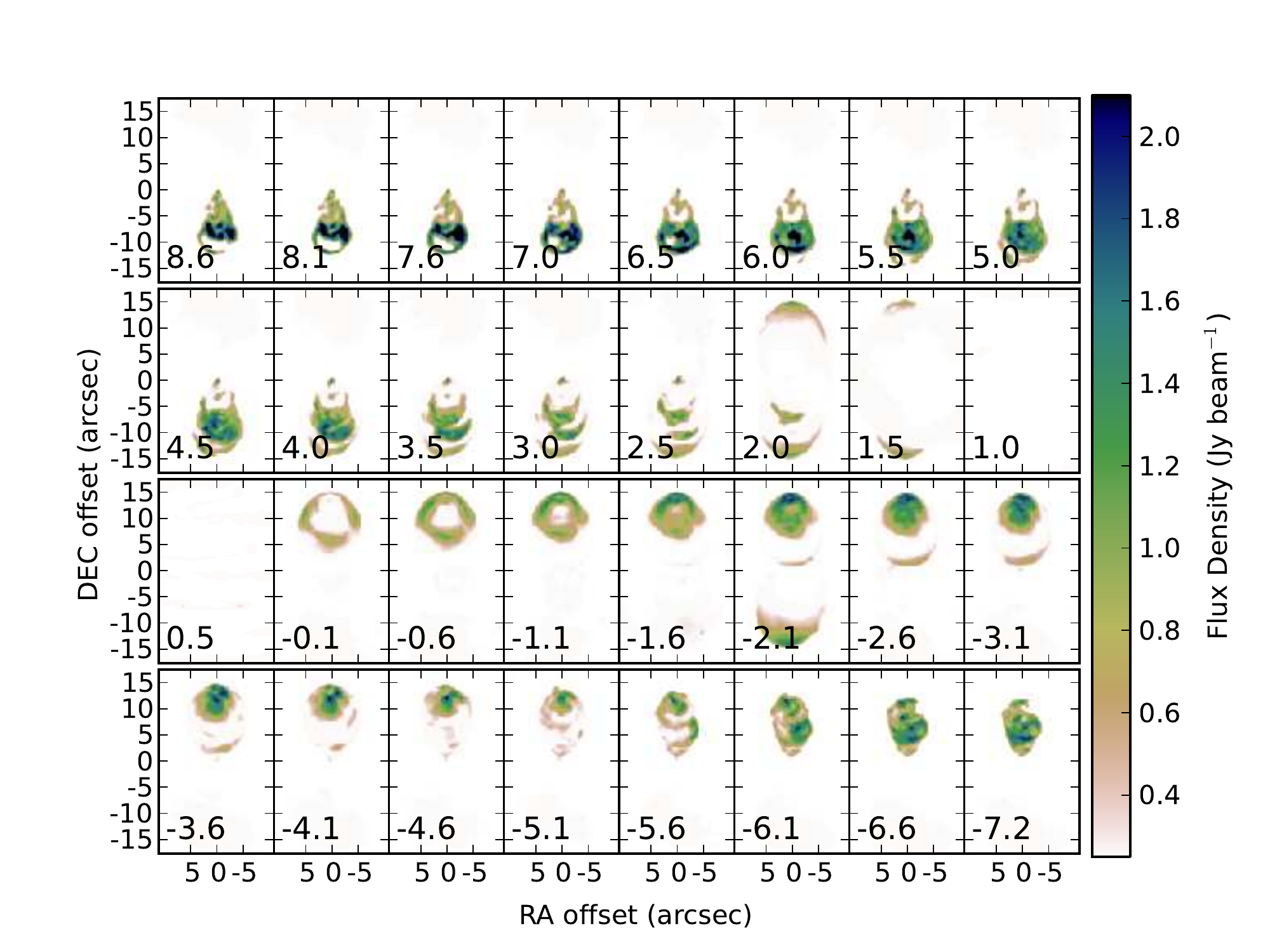}
\caption{First moment map (\emph{left}) and channel maps (\emph{right}) of the CO emission for snapshot Elon-A. Contours range from 0.2 to 9 times the C$^{18}$O peak intensity
(0.83 Jy beam$^{-1}$ for the red and 1.04 Jy beam$^{-1}$ for the blue component).
The number inside each panel of the channel maps is the line-of-sight velocity in km\,s$^{-1}$.}
\label{fig:elona}
\end{figure*}

\begin{figure*}
\includegraphics[width=210pt]{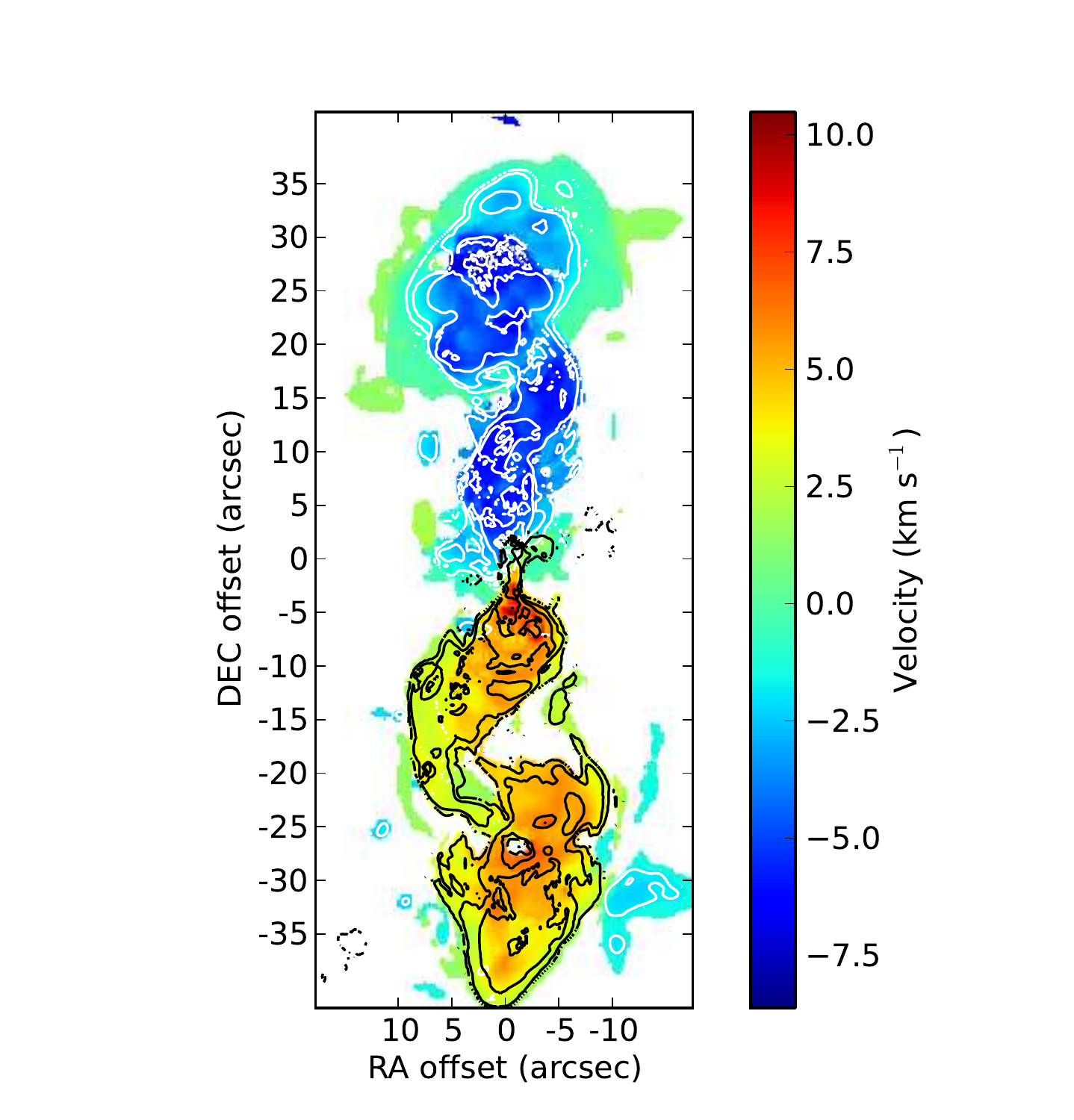}
\includegraphics[width=290pt]{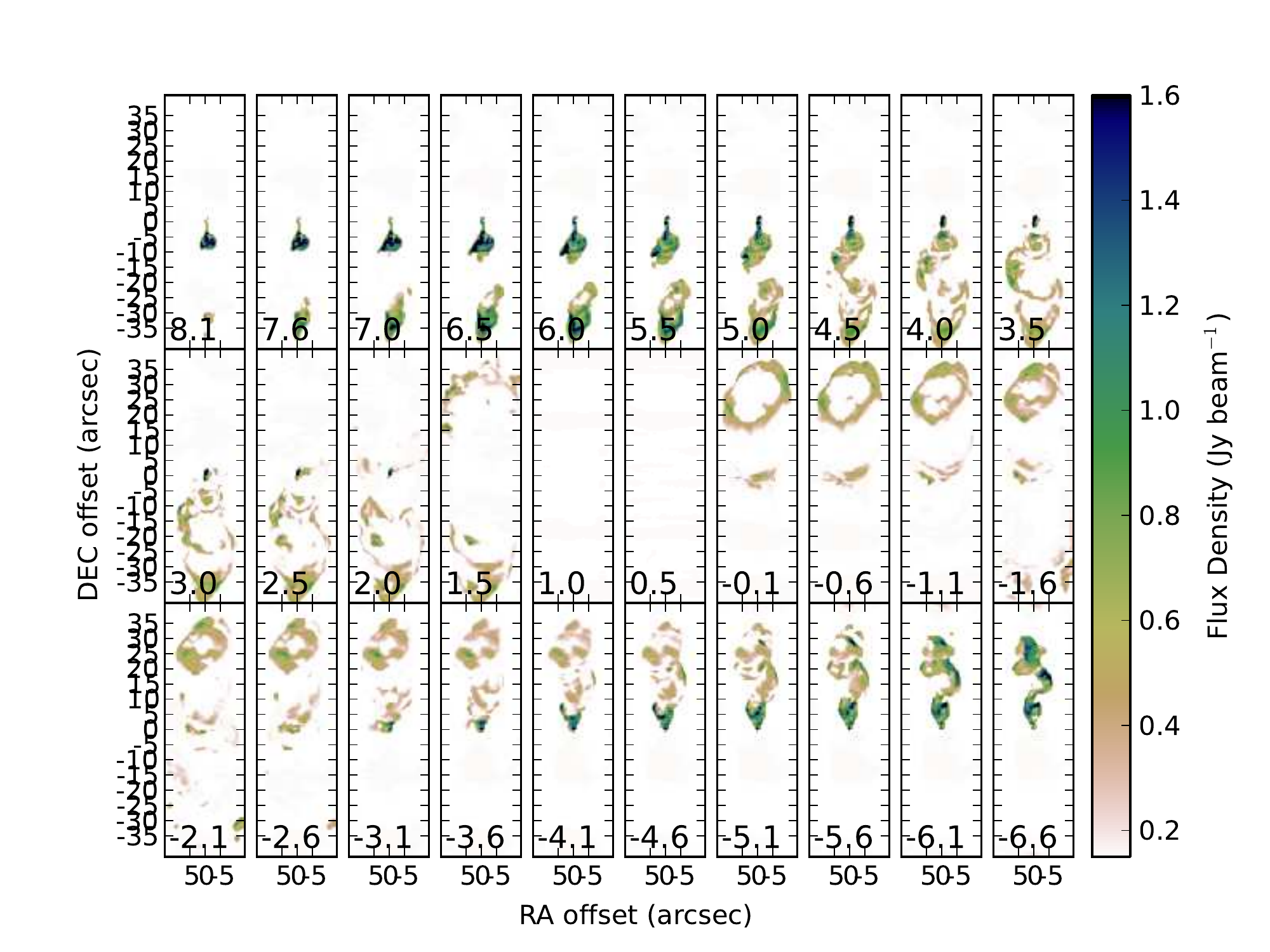}
\caption{First moment map (\emph{left}) and channel maps (\emph{right}) of the CO emission for snapshot Elon-B. Contours range from 0.2 to 9 times the C$^{18}$O peak intensity
(0.33 Jy beam$^{-1}$ for the red and 0.51 Jy beam$^{-1}$ for the blue component). Note that
the spatial scale is different from Figure~\ref{fig:elona}. The number inside each panel of the channel maps is the line-of-sight velocity in km\,s$^{-1}$.}
\label{fig:elonb}
\end{figure*}

\begin{figure*}
\includegraphics[width=210pt]{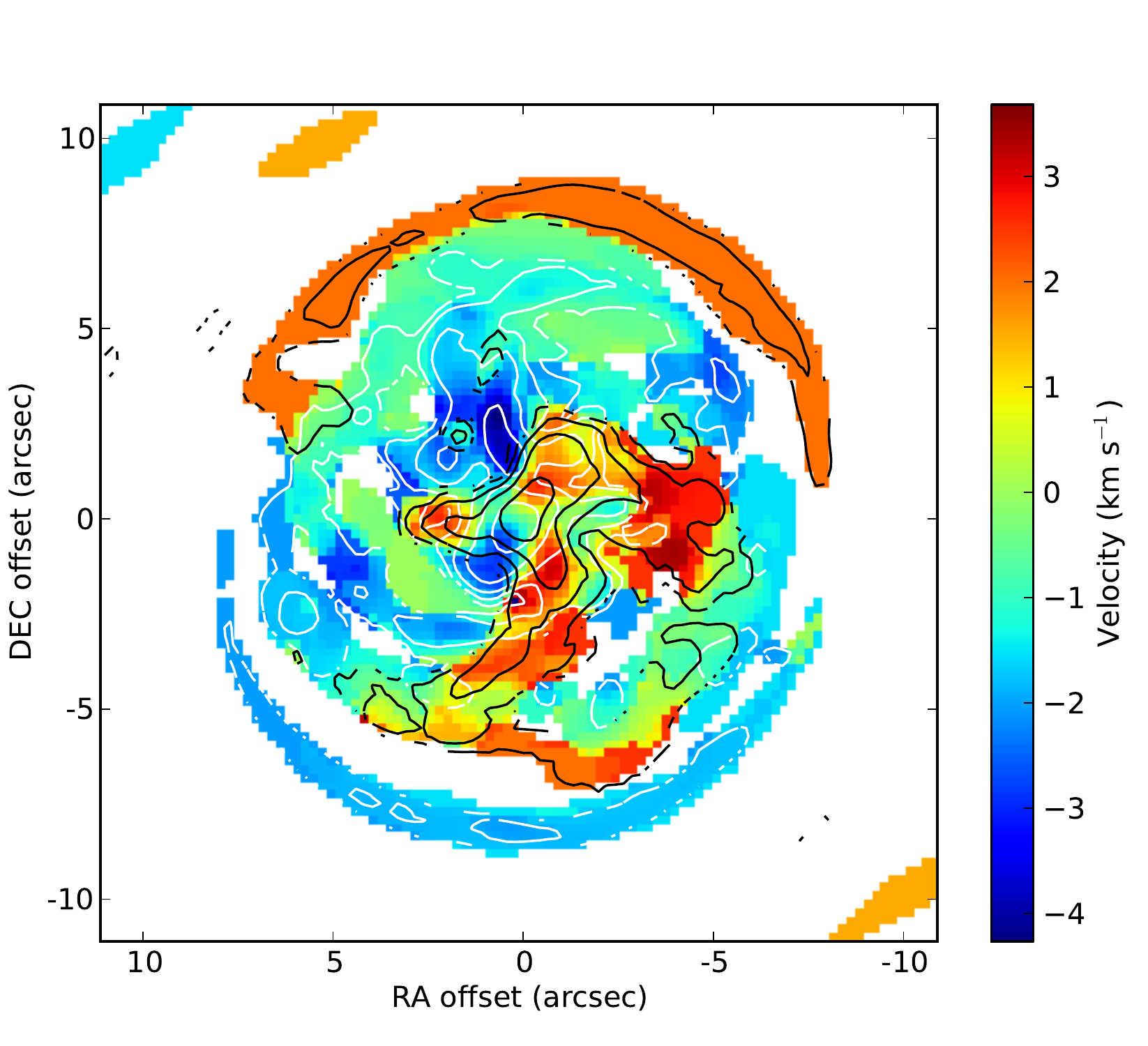}
\includegraphics[width=290pt]{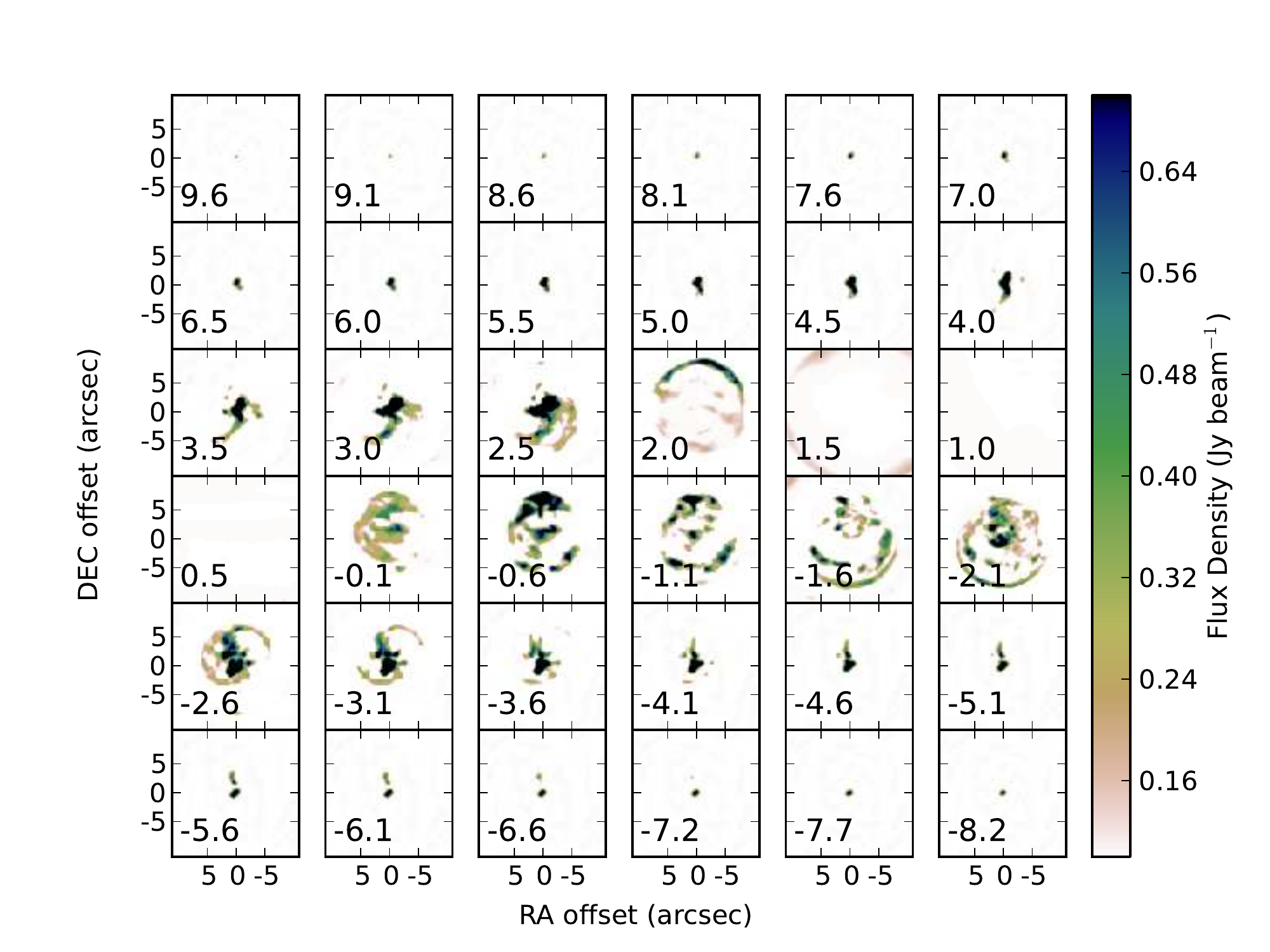}
\caption{First moment map (\emph{left}) and channel maps (\emph{right}) of the CO emission for snapshot Spher. Contours range from 0.2 to 9 times the C$^{18}$O peak intensity
(0.32 Jy beam$^{-1}$ for the red and 0.23 Jy beam$^{-1}$ for the blue component). The
spatial scale is similar to Figure~\ref{fig:elona}. The number inside each panel of the channel maps is the line-of-sight velocity in km\,s$^{-1}$.}
\label{fig:spher}
\end{figure*}

The first moment and channel maps for Elon-A, Elon-B and Spher are shown in Figures~\ref{fig:elona}, \ref{fig:elonb} and
\ref{fig:spher}, respectively. Figure~\ref{fig:elona} illustrates that the outflow velocities closer to the star are generally larger.
This is because the material at a location near the star was launched at a later time than material further away.
At these later times, the star is more massive and the Keplerian velocity greater, increasing the outflow velocity.
In general, the velocities in the first moment map well represent the average velocities in the outflow.

\begin{figure}
\includegraphics[width=260pt]{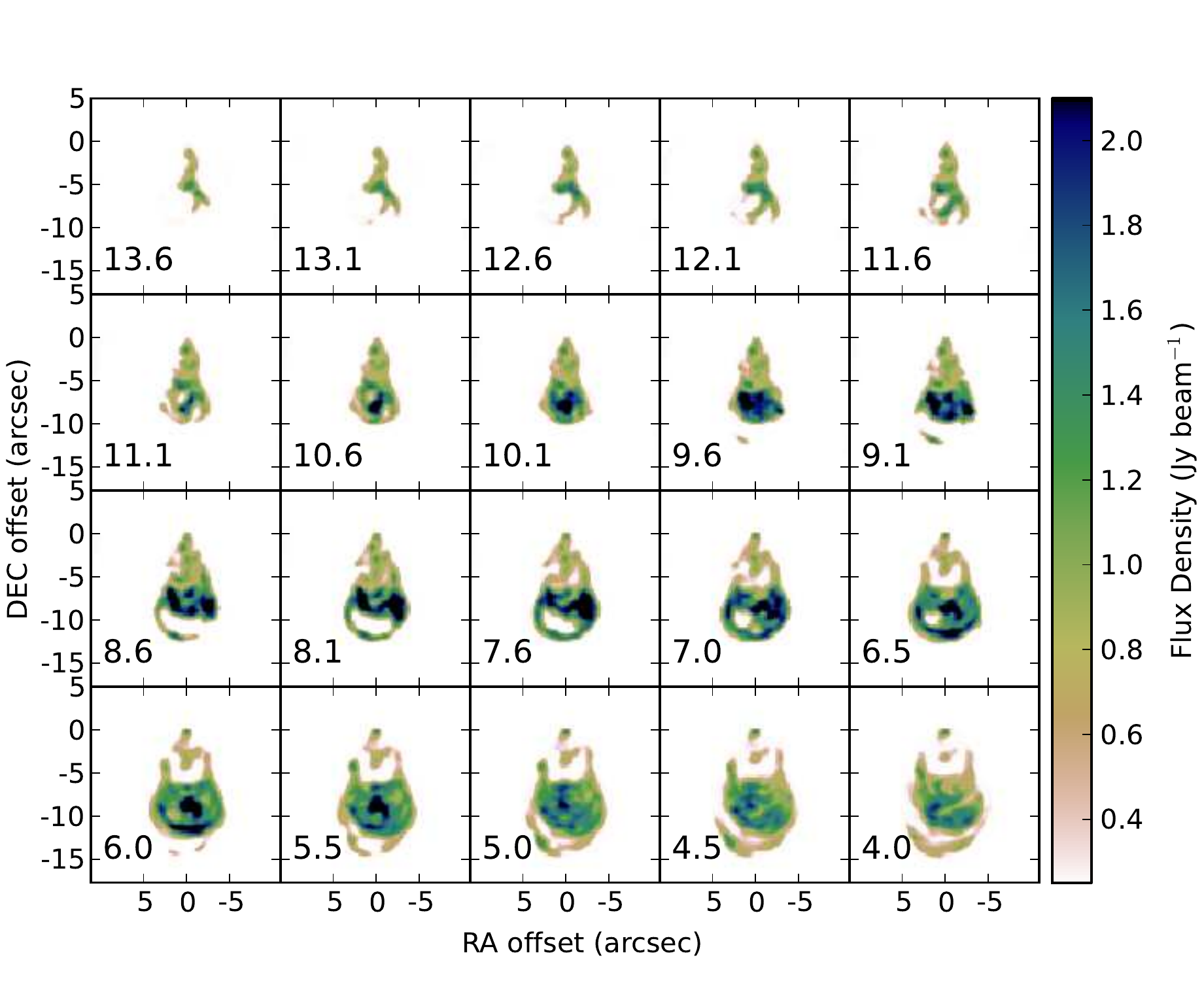}
\caption{Zoom in on the red-shifted lobe of Elon-A, which shows evidence for a tightly wound spiral at high velocities. This helical structure becomes much more pronounced
in the first moment map of Elon-B at later times (see Figure~\ref{fig:elonb}). The number inside each panel of the channel maps is the line-of-sight velocity in km\,s$^{-1}$.}
\label{fig:elona_zoom}
\end{figure}

In Figure~\ref{fig:elonb}, the outflow of Elon-B is shown at a larger scale than Elon-A.
The first moment map displays a very prominent helical structure.
This helix is the result of an MHD instability occuring for axisymmetric jets, the so-called kink
instability~\citep[e.g.][]{ray81,appl92}. The kink instability describes a helical ($m = 1$) displacement of the jet
from the symmetry axis without any distortion of the jet profile. The instability is stabilised by the magnetic field of the jet.
The development of such a kink (or helical) instability over time was studied numerically by~\citet{ouyclapud03} who find a
successive growth over time. Here we find that the instability starts to grow significantly only after the time at which
snapshot Elon-A is taken, which is why it is seen more prominently in Elon-B.
In the maps of Elon-A, there are velocity structures in the individual
channel maps which may be hinting at the existence of a tightly wound rotating structure within the outflow (see Figure \ref{fig:elona_zoom}), however it is
only at later times (in Elon-B), that the helical structure in the outflow becomes identifiable in the first moment map. The bow shock of the outflow in Elon-B
is visible as a low-velocity shell around the tips of the two outflow lobes in Figure~\ref{fig:elonb}.

The velocities and flux densities measured from Spher (Figure~\ref{fig:spher}) are much smaller than those obtained from Elon-A and Elon-B. This is not surprising
since the velocity component along the line of sight is much smaller for the spherical outflow than for the elongated one. Quantitatively, the outflow velocities in Spher
are a factor of a few smaller than the maser spots in W75N \citep{kimkim2013}, on the other hand the size of Spher is also larger by a similar factor.
\citet{seifried12} speculated that the spherical outflow seen in snapshot Spher might be a transient feature because the outflow velocity is so small
that the outflow could fall back onto the disk. With time, a small Keplerian disk around the central star could then build up and make the launching of a collimated
outflow possible. If this scenario is true, then spherical outflows around intermediate-size stars of a size much larger than Spher should not
be observed. However, an outflow similar to the one in W75N has been observed around a massive star in Cepheus A \citep{torrelles01}. A source
in HL Tauri that is likely more evolved than the \citet{seifried12} simulations is surrounded by a spherical bubble as well \citep{welch00}.
These observations might suggest that
spherical outflows can occur repeatedly during the disk evolution, and not only in an initial transient phase.
Since Figure~\ref{fig:spher} does not resemble
a typical protostellar outflow, we think that it might be useful as a reference for observers in case they find a similarly looking object.

The outflow mass and kinematics derived from the simulated observations (see Table~\ref{table:co}) are directly comparable to those derived from the
modelled outflows themselves (see~Table~\ref{table:direct}). The values for mass, momentum and energy are mostly within a factor of two of each other.
The majority of the directly measured values are smaller than the observed ones. The origin of this behaviour is unclear, and we have seen the opposite
trend in previous work \citep{petersetal12}.
The luminosity and mass-loss rate are slightly less accurate because the observationally determined outflow ages are
generally less than half of the true values measured directly from the simulations.
For Elon-A and Elon-B the mass measurements are most consistent, while the measurements of the outflow momentum and energy show larger deviations.
For snapshot Spher, the mass and kinematics derived from the simulated observations are most uncertain because of the spherical outflow morphology.
  Although we have
made every attempt to recover all of the flux in the observed maps and thoroughly cover the uv plane, there may be some missing or enhanced structure
which is biasing our derived masses and kinematics. Uncertainties also arise in the derivation of the optical depth of the lines which could bias our
derivations, which then over- or underestimates the correction factor in Equation~\eqref{eq:tau}.
We used standard observational methods for deriving the outflow masses and kinematics from the simulated observations. That our results are
so strikingly similar to the masses and kinematics derived from the models themselves is a testament to the robustness of the methods used.

\section{Comparison with observations}
\label{sec:compobs}

\begin{figure}
\includegraphics[width=260pt]{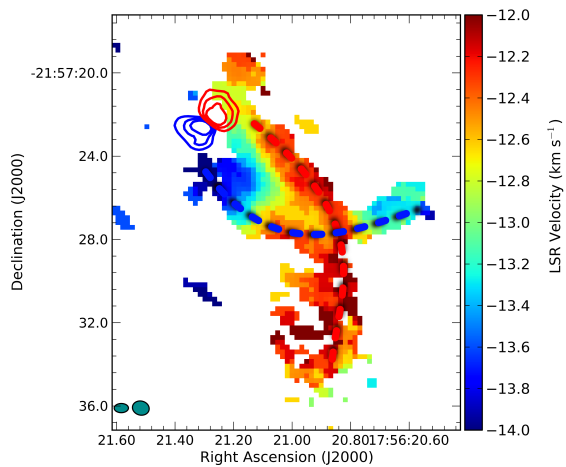}
\caption{First moment CO $J=2-1$ map of HD~163296. The blue and red contours display the 15, 20 and 25 times root-mean-square noise (20~mJy~beam$^{-1}$)
of the blue (-2 to 5.5 km~s$^{-1}$) and red (6 to 12 km~s$^{-1}$) HCO$^+$ $J=4-3$ emission from the disk. The brown contours show 4, 6, and 8 times the
root-mean-square noise in the CO $J=3-2$ emission (2.6 mJy~beam$^{-1}$). The velocities listed in the colour bar are LSR velocities. The rest velocity of the
source is 7 km~s$^{-1}$. The blue and red dashed lines delineate the helical structure mentioned in the text.}
\label{fig:obs}
\end{figure}

The helical structure in snapshot Elon-B can be compared to recent ALMA observations of a disk wind\footnote{Here we use the words ``disk wind'' and ``outflow'' synonymously.
The former term is used in \citet{klaassen2013} because the outflow can be traced unequivocally to the disk from which it is launched.} around the Herbig Ae~star HD~163296 \citep{klaassen2013}.
We show the first moment CO map of HD~163296 in Figure~\ref{fig:obs}. For reference, the HCO$^+$ emission from the disk is also shown as blue and red contours.
HD~163296 is located at a distance of 122~pc, and therefore the angular scales in Figure~\ref{fig:obs} are directly comparable to those
in Figure~\ref{fig:elonb}. One can see that the spatial scale of the observed helically
twisted arcs fits very nicely to our model prediction. In fact, the kink of the helix is at the same distance from the central star.
The velocities of the HD 163296 wind are, however, much faster than those seen in Elon-B (averaging at 18.6~km~s$^{-1}$ from the source rest velocity).
The limits of the observations mean that the large-scale morphology of this wind has yet to be observed, but it is possible that
the observed structures in HD~163296 do wind up to a spiral on larger scales.
Even though HD~163296 is about 4~Myr old, we compare it to Elon-B (where the
protostar is about 10 kyr old) because observations of HD~163296 show for the
first time the corkscrew structure that our simulations predict.

If we degrade the resolution and sensitivity in the maps of Elon-B, we would still be able to see velocity gradients within the outflow, such as in Elon-A. \citet{pech12}
have reported velocity gradients in CO observations of HH 797 with the SMA. They report velocity differences of 2~km~s$^{-1}$ over distances of 1000~AU, which are roughly
consistent with the velocity gradient in Elon-A. ALMA, with more than 10 times the linear resolution of the SMA, will likely see more than a velocity gradient in each
lobe of HH 797, but the precessing gas itself.

\begin{figure}
\centerline{\includegraphics[height=200pt]{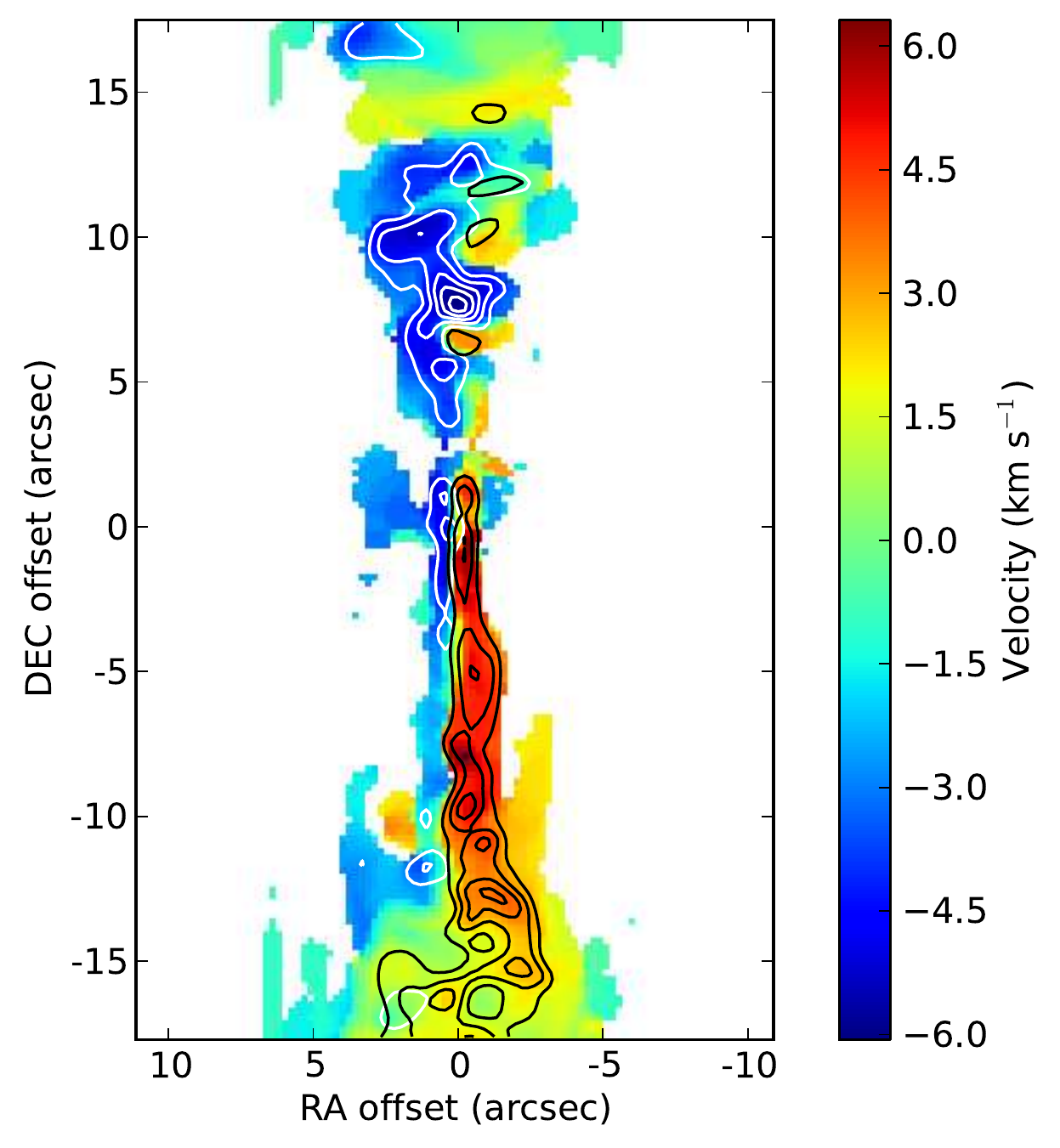}}
\caption{Edge-on first moment map of the CO emission for snapshot Elon-A. Contours range from 0.2 to 9 times the C$^{18}$O peak intensity.}
\label{fig:edgeona}
\end{figure}

\begin{figure}
\centerline{\includegraphics[height=200pt]{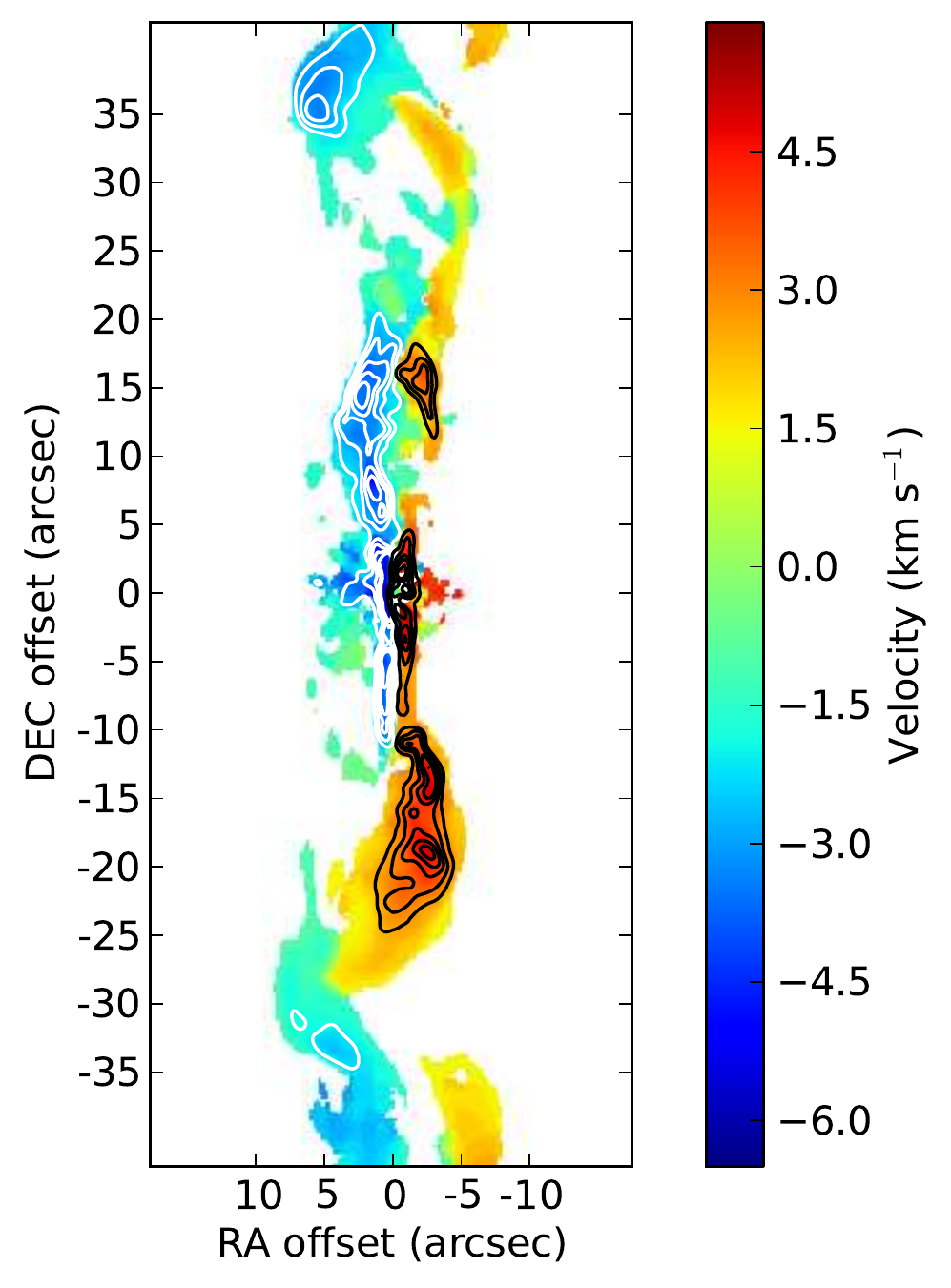}}
\caption{Edge-on first moment map of the CO emission for snapshot Elon-B. Contours range from 0.2 to 9 times the C$^{18}$O peak intensity.}
\label{fig:edgeonb}
\end{figure}

The rotation of the protostellar outflow can be seen at larger inclination angles.
Figure~\ref{fig:edgeona} and \ref{fig:edgeonb} show CO observations of Elon-A and Elon-B, respectively, at an inclination of 80$^\circ$.
We have chosen a small deviation from 90$^\circ$ since observations exactly edge-on are very unlikely.
These maps show velocity gradients across the outflow that are clear evidence for outflow rotation. Additionaly, there are velocity gradients
along the outflow axis. These secondary gradients are caused by the growth of the Keplerian velocity with time as well as the complex gas motion along the helical structure,
exactly the same reasons as for the 30$^\circ$ maps.
Similar gradients have been found in CO observations of an outflow in CB~26 with the PdBI by \citet{launhardt09}. 

In their study of 16 Class 0 sources, \citet{tobin11} found that in addition to the outflow motions from these sources, in more than half of them (11), there were extra
velocity gradients more than 45$^\circ$ from the outflow direction. They suggested that this was either due to contamination by infall, or due to rotation
of the outflowing gas itself. However, our models Elon-A and Elon-B also show significant velocity gradients not aligned with the outflow axis. These gradients
are created by the complex dynamics of the outflowing gas and by shocked gas in the outflow, not by infall.

In HD~163296, the outflow can be traced to the disk from which it is launched. There are other recent ALMA observations of protostellar outflows in which this connection has not been made.
\citet{zapata12o} have imaged an outflow around a young massive star in Orion-KL in SiO. Since SiO is a shock tracer and the observations do not have
many resolution elements across the outflow, we cannot expect to see a helical structure. \citet{merello13} have reported ALMA observations of one of the most
energetic and luminous outflows in the Milky Way, G331.512-0.103. This outflow is also not well resolved, and our outflows are certainly much weaker.
\citet{arce13} have presented ALMA CO observations of the HH 46/47 molecular outflow. This outflow is much smaller in mass, momentum and energy than any
of our outflows, and the associated star appears to be a little bit smaller, too. However, there appears to be no evidence for a helical structure
in the outflow.

\section{Conclusions}
\label{sec:concl}

We have presented synthetic ALMA observations of intermediate-mass outflows, derived from the self-consistent magnetohydrodynamic protostellar collapse calculations of
\citet{seifried12}. We find generally good agreement between outflow properties measured from CO lines and the simulation data.
Elongated outflows are much easier to detect than spherical outflows because of their larger line-of-sight velocities. However, the dearth of evidence for spherical
outflows could also be explained on statistical grounds if they are just transient objects. More observations of spherical outflows and simulations that
follow the disk and outflow evolution for a longer period are necessary to settle the question.
Edge-on views of the elongated
outflows show velocity gradients consistent with observations.
We find a helical structure in the CO maps that is caused by an instability during the outflow launching.
This helix is already present in the early outflow phases ($\sim$5000~yr) but becomes very prominent at later times ($\sim$10000~yr). We speculate
that the recent observation of a disk wind in HD~163296 is the first instance of the detection of such a helix, which needs to be backed up by follow-up studies.

\section*{Acknowledgements}

We thank the anonymous referee for useful comments that helped to improve the paper.
T.P. acknowledges financial support through SNF grant 200020\textunderscore 137896 and a Forschungskredit of the University of Z\"{u}rich, grant no. FK-13-112.
D.S. acknowledges funding from the {\em Deutsche Forschungsgemeinschaft} DFG via the grant BA 3706/1-3 within the Priority Program SPP 1573 {\em Physics of the Interstellar Medium}.
R.B. acknowledges funding from the DFG via the grant BA 3706/1-1.
R.S.K. acknowledges funding from the DFG via grants KL~1358/14-1 as part of the SPP 1573 as well as via the Collaborative Research Project
SBB~811 {\em The Milky Way System} in subprojects B1, B2, and B4.
The FLASH code was in part developed by the DOE-supported Alliances Center for Astrophysical Thermonuclear Flashes (ASCI) at the University of Chicago.


\begin{thebibliography}{66}
\expandafter\ifx\csname natexlab\endcsname\relax\def\natexlab#1{#1}\fi

\bibitem[{Appl \& Camenzind(1992)}]{appl92}
Appl, S., \& Camenzind, M. 1992, \aap, 256, 354

\bibitem[{Arce {et~al.}(2013)Arce, Mardones, Corder, Garay, Noriega-Crespo, \&
  Raga}]{arce13}
Arce, H.~G., Mardones, D., Corder, S.~A., {et~al.} 2013, \apj, 774, 39

\bibitem[{Arce {et~al.}(2007)Arce, Shepherd, Gueth, Lee, Bachiller, Rosen, \&
  Beuther}]{arceetal07}
Arce, H.~G., Shepherd, D., Gueth, F., {et~al.} 2007, in {Protostars and Planets
  V}, ed. B.~Reipurth, D.~Jewitt, \& K.~Keil ({Tucson: The University of
  Arizona Press}), 245

\bibitem[{Bachiller(1996)}]{bachiller96}
Bachiller, R. 1996, \araa, 34, 111

\bibitem[{Bally(2007)}]{bally07}
Bally, J. 2007, \apss, 311, 15

\bibitem[{Bally(2008)}]{bally08}
Bally, J. 2008, in {Astronomical Society of the Pacific Conference Series},
  Vol. 387, {Massive Star Formation: Observations Confront Theory}, ed.
  H.~Beuther, H.~Linz, \& T.~Henning, 158

\bibitem[{Bally {et~al.}(2007)Bally, Reipurth, \& Davis}]{baletal07}
Bally, J., Reipurth, B., \& Davis, C.~J. 2007, in {Protostars and Planets V},
  ed. B.~Reipurth, D.~Jewitt, \& K.~Keil ({Tucson: The University of Arizona
  Press}), 215

\bibitem[{Beuther {et~al.}(2004)Beuther, Schilke, \& Gueth}]{beuthetal04a}
Beuther, H., Schilke, P., \& Gueth, F. 2004, \apj, 608, 330

\bibitem[{Beuther {et~al.}(2002{\natexlab{a}})Beuther, Schilke, Gueth,
  McCaughrean, Andersen, Sridharan, \& Menten}]{beutheretal02b}
Beuther, H., Schilke, P., Gueth, F., {et~al.} 2002{\natexlab{a}}, \aap, 387,
  931

\bibitem[{Beuther {et~al.}(2002{\natexlab{b}})Beuther, Schilke, Sridharan,
  Menten, Walmsley, \& Wyrowski}]{beutheretal02}
Beuther, H., Schilke, P., Sridharan, T.~K., {et~al.} 2002{\natexlab{b}}, \aap,
  383, 892

\bibitem[{Beuther \& Shepherd(2005)}]{beuthshep05}
Beuther, H., \& Shepherd, D. 2005, in {Cores to Clusters: Star Formation with
  Next Generation Telescopes}, ed. M.~S.~N. Kumar, M.~Tafalla, \& P.~Caselli
  ({Springer-Verlag}), 105

\bibitem[{Blandford \& Payne(1982)}]{blandfordpayne82}
Blandford, R.~D., \& Payne, D.~G. 1982, \mnras, 199, 883

\bibitem[{Bouchut {et~al.}(2007)Bouchut, Klingenberg, \& Waagan}]{bouchut07}
Bouchut, F., Klingenberg, C., \& Waagan, K. 2007, Numerische Mathematik, 108, 7

\bibitem[{Cabrit \& Andr{\'e}(1991)}]{caband91}
Cabrit, S., \& Andr{\'e}, P. 1991, \apj, 379, L25

\bibitem[{Cabrit \& Bertout(1986)}]{cabber86}
Cabrit, S., \& Bertout, C. 1986, \apj, 307, 313

\bibitem[{Choi {et~al.}(1993)Choi, Evans, \& Jaffe}]{choi93}
Choi, M., Evans, II, N.~J., \& Jaffe, D.~T. 1993, \apj, 417, 624

\bibitem[{Federrath {et~al.}(2010)Federrath, Banerjee, Clark, \&
  Klessen}]{federrathetal10}
Federrath, C., Banerjee, R., Clark, P.~C., \& Klessen, R.~S. 2010, \apj, 713,
  269

\bibitem[{Fryxell {et~al.}(2000)Fryxell, Olson, Ricker, Timmes, Zingale, Lamb,
  MacNeice, Rosner, Truran, \& Tufo}]{fryxell00}
Fryxell, B., Olson, K., Ricker, P., {et~al.} 2000, \apjs, 131, 273

\bibitem[{Girart {et~al.}(2013)Girart, Frau, Zhang, Koch, Qiu, Tang, Lai, \&
  Ho}]{giratetal2013}
Girart, J.~M., Frau, P., Zhang, Q., {et~al.} 2013, \apj, 772, 69

\bibitem[{Girichidis {et~al.}(2012)Girichidis, Federrath, Banerjee, \&
  Klessen}]{girietal12}
Girichidis, P., Federrath, C., Banerjee, R., \& Klessen, R.~S. 2012, \mnras,
  420, 613

\bibitem[{Hennebelle \& Ciardi(2009)}]{hencia09}
Hennebelle, P., \& Ciardi, A. 2009, \aap, 506, L29

\bibitem[{Hennebelle \& Teyssier(2008)}]{hentey08}
Hennebelle, P., \& Teyssier, R. 2008, \aap, 477, 25

\bibitem[{Kim {et~al.}(2013)Kim, Kim, Kurayama, Honma, Sasao, Surcis,
  {Cant{\'o}}, Torrelles, \& Kim}]{kimkim2013}
Kim, J.-S., Kim, S.-W., Kurayama, T., {et~al.} 2013, \apj, 767, 86

\bibitem[{Klaassen {et~al.}(2013{\natexlab{a}})Klaassen, Galv{\'a}n-Madrid,
  Peters, Longmore, \& Maercker}]{klaassen2013b}
Klaassen, P.~D., Galv{\'a}n-Madrid, R., Peters, T., Longmore, S.~N., \&
  Maercker, M. 2013{\natexlab{a}}, \aap, 556, A107

\bibitem[{Klaassen {et~al.}(2013{\natexlab{b}})Klaassen, Juhasz, Mathews,
  Mottram, {De Gregorio-Monsalvo}, {van Dishoeck}, Takahashi, Akiyama,
  Chapillon, Espada, Hales, Hogerheijde, Rawlings, Schmalzl, \&
  Testi}]{klaassen2013}
Klaassen, P.~D., Juhasz, A., Mathews, G.~S., {et~al.} 2013{\natexlab{b}}, \aap,
  555, A73

\bibitem[{Krumholz {et~al.}(2009)Krumholz, Klein, McKee, Offner, \&
  Cunningham}]{krumholzetal09}
Krumholz, M.~R., Klein, R.~I., McKee, C.~F., Offner, S.~S.~R., \& Cunningham,
  A.~J. 2009, {Science}, 323, 754

\bibitem[{Kuiper {et~al.}(2012)Kuiper, Klahr, Beuther, \&
  Henning}]{kuiperetal12}
Kuiper, R., Klahr, H., Beuther, H., \& Henning, T. 2012, \aap, 537, A122

\bibitem[{Launhardt {et~al.}(2009)Launhardt, Pavlyuchenkov, Gueth, Chen,
  Dutrey, Guilloteau, Henning, {Pi{\'e}tu}, Schreyer, \& Semenov}]{launhardt09}
Launhardt, R., Pavlyuchenkov, Y., Gueth, F., {et~al.} 2009, \aap, 494, 147

\bibitem[{Lynden-Bell(1996)}]{lynbel96}
Lynden-Bell, D. 1996, \mnras, 279, 389

\bibitem[{Lynden-Bell(2003)}]{lynbel03}
---. 2003, \mnras, 341, 1360

\bibitem[{McMullin {et~al.}(2007)McMullin, Waters, Schiebel, Young, \&
  Golap}]{mcmullin07}
McMullin, J.~P., Waters, B., Schiebel, D., Young, W., \& Golap, K. 2007, in
  {Astronomical Data Analysis Software and Systems XVI}, ed. R.~A. Shaw,
  F.~Hill, \& D.~J. Bell ({San Francisco: ASP}), 127

\bibitem[{Merello {et~al.}(2013)Merello, Bronfman, Garay, Lo, Evans, Nyman,
  Cort{\'e}s, \& Cunningham}]{merello13}
Merello, M., Bronfman, L., Garay, G., {et~al.} 2013, \apj, 774, L7

\bibitem[{Mouschovias \& Paleologou(1980)}]{moupal80}
Mouschovias, T.~C., \& Paleologou, E.~V. 1980, \apj, 237, 877

\bibitem[{Myers {et~al.}(2013)Myers, McKee, Cunningham, Klein, \&
  Krumholz}]{myers13}
Myers, A.~T., McKee, C.~F., Cunningham, A.~J., Klein, R.~I., \& Krumholz, M.~R.
  2013, \apj, 766, 97

\bibitem[{Ouyed {et~al.}(2003)Ouyed, Clarke, \& Pudritz}]{ouyclapud03}
Ouyed, R., Clarke, D.~A., \& Pudritz, R.~E. 2003, \apj, 582, 292

\bibitem[{Pech {et~al.}(2012)Pech, Zapata, Loinard, \&
  {Rodr{\'{\i}}guez}}]{pech12}
Pech, G., Zapata, L.~A., Loinard, L., \& {Rodr{\'{\i}}guez}, L.~F. 2012, \apj,
  751, 78

\bibitem[{Pelletier \& Pudritz(1992)}]{pelput92}
Pelletier, G., \& Pudritz, R.~E. 1992, \apj, 394, 117

\bibitem[{Peters {et~al.}(2011)Peters, Banerjee, Klessen, \&
  Mac~Low}]{petersetal11a}
Peters, T., Banerjee, R., Klessen, R.~S., \& Mac~Low, M.-M. 2011, \apj, 729, 72

\bibitem[{Peters {et~al.}(2010{\natexlab{a}})Peters, Banerjee, Klessen,
  Mac~Low, Galv{\'a}n-Madrid, \& Keto}]{petersetal10a}
Peters, T., Banerjee, R., Klessen, R.~S., {et~al.} 2010{\natexlab{a}}, \apj,
  711, 1017

\bibitem[{Peters {et~al.}(2012)Peters, Klaassen, Mac~Low, Klessen, \&
  Banerjee}]{petersetal12}
Peters, T., Klaassen, P.~D., Mac~Low, M.-M., Klessen, R.~S., \& Banerjee, R.
  2012, \apj, 760, 91

\bibitem[{Peters {et~al.}(2010{\natexlab{b}})Peters, Klessen, Mac~Low, \&
  Banerjee}]{petersetal10c}
Peters, T., Klessen, R.~S., Mac~Low, M.-M., \& Banerjee, R. 2010{\natexlab{b}},
  \apj, 725, 134

\bibitem[{Pudritz \& Norman(1983)}]{pudrnorm83}
Pudritz, R.~E., \& Norman, C.~A. 1983, \apj, 274, 677

\bibitem[{Ray {et~al.}(2007)Ray, Dougados, Bacciotti, Eisl{\"o}ffel, \&
  Chrysostomou}]{rayetal07}
Ray, T., Dougados, C., Bacciotti, F., Eisl{\"o}ffel, J., \& Chrysostomou, A.
  2007, in {Protostars and Planets V}, ed. B.~Reipurth, D.~Jewitt, \& K.~Keil
  ({Tucson: The University of Arizona Press}), 231

\bibitem[{Ray(1981)}]{ray81}
Ray, T.~P. 1981, \mnras, 196, 195

\bibitem[{Reipurth \& Bally(2001)}]{reibal01}
Reipurth, B., \& Bally, J. 2001, \araa, 39, 403

\bibitem[{Ren {et~al.}(2011)Ren, Liu, Wu, \& Li}]{renetal11}
Ren, J.~Z., Liu, T., Wu, Y., \& Li, L. 2011, \mnras, 415, L49

\bibitem[{Ridge \& Moore(2001)}]{ridmoo01}
Ridge, N.~A., \& Moore, T.~J.~T. 2001, \aap, 378, 495

\bibitem[{{Santos-Lima} {et~al.}(2012){Santos-Lima}, {de Gouveia Dal Pino}, \&
  Lazarian}]{santoslima12}
{Santos-Lima}, R., {de Gouveia Dal Pino}, E.~M., \& Lazarian, A. 2012, \apj,
  747, 21

\bibitem[{Sch{\"o}ier {et~al.}(2005)Sch{\"o}ier, {van der Tak}, {van Dishoeck},
  \& Black}]{schoeieretal05}
Sch{\"o}ier, F.~L., {van der Tak}, F.~F.~S., {van Dishoeck}, E.~F., \& Black,
  J.~H. 2005, \aap, 432, 369

\bibitem[{Seifried {et~al.}(2011)Seifried, Banerjee, Klessen, Duffin, \&
  Pudritz}]{seifetal11}
Seifried, D., Banerjee, R., Klessen, R.~S., Duffin, D., \& Pudritz, R.~E. 2011,
  \mnras, 417, 1054

\bibitem[{Seifried {et~al.}(2012{\natexlab{a}})Seifried, Banerjee, Pudritz, \&
  Klessen}]{seifried12a}
Seifried, D., Banerjee, R., Pudritz, R.~E., \& Klessen, R.~S.
  2012{\natexlab{a}}, \mnras, 423, L40

\bibitem[{Seifried {et~al.}(2013)Seifried, Banerjee, Pudritz, \&
  Klessen}]{seifried13}
---. 2013, \mnras, 432, 3320

\bibitem[{Seifried {et~al.}(2012{\natexlab{b}})Seifried, Pudritz, Banerjee,
  Duffin, \& Klessen}]{seifried12}
Seifried, D., Pudritz, R.~E., Banerjee, R., Duffin, D., \& Klessen, R.~S.
  2012{\natexlab{b}}, \mnras, 422, 347

\bibitem[{Shepherd(2005)}]{shepherd05}
Shepherd, D. 2005, in {Massive star birth: A crossroads of Astrophysics}, ed.
  R.~Cesaroni, M.~Felli, E.~Churchwell, \& M.~Walmsley ({Cambridge University
  Press}), 237

\bibitem[{Shi {et~al.}(2010)Shi, Zhao, \& Han}]{shietal10}
Shi, H., Zhao, J.-H., \& Han, J.~L. 2010, \apj, 718, L181

\bibitem[{Staff {et~al.}(2010)Staff, Niebergal, Ouyed, Pudritz, \&
  Cai}]{staff10}
Staff, J.~E., Niebergal, B.~P., Ouyed, R., Pudritz, R.~E., \& Cai, K. 2010,
  \apj, 722, 1325

\bibitem[{Surcis {et~al.}(2011)Surcis, Vlemmings, Curiel, {Hutawarakorn
  Kramer}, Torrelles, \& Sarma}]{surcisetal11}
Surcis, G., Vlemmings, W.~H.~T., Curiel, S., {et~al.} 2011, \aap, 527, A48

\bibitem[{Tobin {et~al.}(2011)Tobin, Hartmann, Chiang, Looney, Bergin,
  Chandler, {Masqu{\'e}}, Maret, \& Heitsch}]{tobin11}
Tobin, J.~J., Hartmann, L., Chiang, H.-F., {et~al.} 2011, \apj, 740, 45

\bibitem[{Torrelles {et~al.}(2001)Torrelles, Patel, G{\'o}mez, Ho,
  Rodr{\'{\i}}guez, Anglada, Garay, Greenhill, Curiel, \&
  Cant{\'o}}]{torrelles01}
Torrelles, J.~M., Patel, N.~A., G{\'o}mez, J.~F., {et~al.} 2001, \nat, 411, 277

\bibitem[{Torrelles {et~al.}(2003)Torrelles, Patel, Anglada, {G{\'o}mez}, Ho,
  Lara, Alberdi, {Cant{\'o}}, Curiel, Garay, \&
  {Rodr{\'{\i}}guez}}]{torrelles2003}
Torrelles, J.~M., Patel, N.~A., Anglada, G., {et~al.} 2003, \apj, 598, L115

\bibitem[{Wang {et~al.}(2011)Wang, Beuther, Bik, Vasyunina, Jiang, Puga, Linz,
  {Rod{\'o}n}, Henning, \& Tamura}]{wangetal11}
Wang, Y., Beuther, H., Bik, A., {et~al.} 2011, \aap, 527, A32

\bibitem[{Welch {et~al.}(2000)Welch, Hartmann, Helfer, \&
  Brice{\~n}o}]{welch00}
Welch, W.~J., Hartmann, L., Helfer, T., \& Brice{\~n}o, C. 2000, \apj, 540, 362

\bibitem[{Wilson \& Rood(1994)}]{wilroo94}
Wilson, T.~L., \& Rood, R.~T. 1994, \araa, 32, 191

\bibitem[{Wu {et~al.}(2004)Wu, Wei, Zhao, Shi, Yu, Qin, \& Huang}]{wuetal04}
Wu, Y., Wei, Y., Zhao, M., {et~al.} 2004, \aap, 426, 503

\bibitem[{Zapata {et~al.}(2012)Zapata, Rodr{\'{\i}}guez, Schmid-Burgk, Loinard,
  Menten, \& Curiel}]{zapata12o}
Zapata, L.~A., Rodr{\'{\i}}guez, L.~F., Schmid-Burgk, J., {et~al.} 2012, \apj,
  754, L17

\bibitem[{Zhang {et~al.}(2005)Zhang, Hunter, Brand, Sridharan, Cesaroni,
  Molinari, Wang, \& Kramer}]{zhangetal05}
Zhang, Q., Hunter, T.~R., Brand, J., {et~al.} 2005, \apj, 625, 864

\end{thebibliography}
\end{document}